# A Distributed IP-based telecommunication system using SIP

Carlton Andre Thompson[1], Haniph A. Latchman[2], Nathan Angelacos[3], Bharath Kumar Pareek[4]

Department of Electrical and Computer Engineering, University of Florida, Gainesville, Florida

## ABSTRACT

*Voice over Internet Protocol (VoIP) technologies are integral to modern telecommunications because of their advanced features, flexibility, and economic benefits. Internet Service Providers initially promoted these technologies by providing low cost local and international calling. At present, there is also a great deal of interest in using IP-based technologies to replace traditional small and large office telephone systems that use traditional PBX's (Private Branch eXchange). Unfortunately, the large majority of the emerging VoIP based office telephone systems have followed the centralized design of traditional public and private telephone systems in which all the intelligence in the system is at the core, with quite expensive hardware and software components and appropriate redundancy for adequate levels of reliability. In this paper, it is argued that a centralized model for an IP-based telecommunications system fails to exploit the full capabilities of Internet-inspired communications and that, very simple, inexpensive, elegant and flexible solutions are possible by deliberately avoiding the centralized approach. This paper describes the design, philosophy and implementation of a prototype for a fully distributed IP-based Telecommunication System (IPTS) that provides the essential feature set for office and home telecommunications, including IP-based long-distance and local calling, and with the support for video as well as data and text. The prototype system was implemented with an Internet-inspired distributed design using open source software, with appropriate customizations and configurations.*

## 1. INTRODUCTION

In recent years there has been a push by telecommunications service providers to move to IP-based telephony. The two major application domains that have been addressed in the evolution of VoIP technologies are (i) VoIP for telephone call cost reduction and (ii) VoIP for in-office communication. We will briefly examine each of these areas.

### 1.1. VoIP for Telephone Cost Reduction

Providers, such as Vonage and magicJack, have helped to popularize the idea of 'Voice over IP' (VoIP) [1] [2]. Such VoIP telephony has been made possible via an Internet connection by using a VoIP phone adapter, (Figure 1) or a computer application. This allows the connection to an Internet Telephone Service Provider (ITSP) to make free or low cost VoIP calls to the PSTN worldwide. Such systems work by converting the analog signal to a digital format that is sent over the Internet. A VoIP user desiring to call regular PSTN numbers, must first establish an account with an ITSP which then enables VoIP calls via their analog telephone, computer, or mobile device software. VoIP users are able to make free calls to other users of the ITSP system and local, long distance, and international calls to PSTN numbers for a fraction of what it costs via traditional Public Switched Telephone Network (PSTN). For example, at the time of this writing, Vonage offers long distance calls to some countries for as little as $0.05 per minute (call to Belize) [1]. This is in contrast to the cost of $0.48 per minute using the PSTN or $0.55 using a cellphone [3]. This is nearly a 10 times cost reduction by using VoIP versus the PSTN or cellphone. The ITSP typically has a contract with the destination country or intermediary PSTN provider. It should be noted however, in some countries VoIP devices are considered illegal due to the issue of 'revenue-bypass', with outbound international calling being the most sensitive [4]. The tremendous cost savings in using VoIP for local, long distance, and international calling calls result from using the Internet and packet switching to replace the sequence of traditional telephone circuits made up of central office, tandem switches, toll switches, and a variety of circuit based telecommunications links[5]

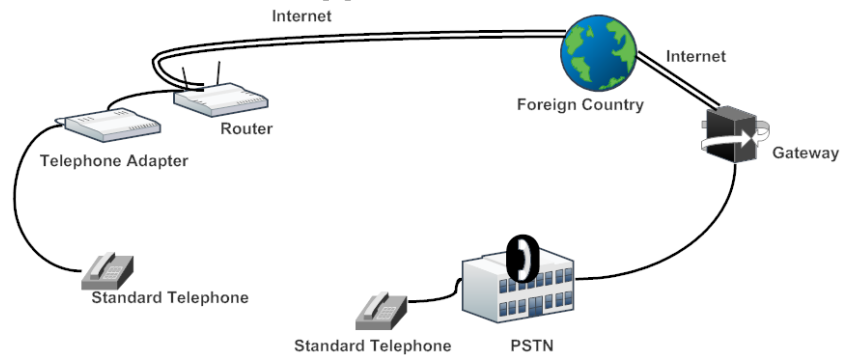

Figure 1. VoIP calling using adapter

## 1.2. VoIP for In-Office Communication

The same IP-based technologies may also be used to reap tremendous cost, performance, and service benefits for in-office communications by replacing the central PBX with an IP-based system for Telecommunications. The traditional PBX centralized design uses component redundancy to increase reliability, Figure 2[6]. The major components (CPU, the Memory, the Switching Matrix and the Power supply), have failover backups in the case of an outage. It must be noted, however, that there is no redundancy at the line and trunk interface level. Thus, in case of line or trunk interface failure, users will experience service disruptions.

Though highly reliable, traditional PBXs for in-office communication are based on a centralized design that goes back to the early days of telephony, with all the intelligence of the telephone system located in the central PBX of the switching system. This design is inherently costly to purchase, maintain, operate and expand. In addition, it also promotes vendor lock-in, where new features are only available only when the vendor supplies them, in a proprietary way. These factors have motivated a great deal of interest in migrating to an open IP-based system using the Session Initiation Protocol (SIP) which supports in-office telephony and at the same time allows access to the cost savings in VoIP calling. In contrast to proprietary PBX's, as an IETF open standard, customer driven SIP customizations are now possible and cost-effective, and feature prominently in the feature set implementations of the IP-based Telecommunication System (IPTS) described in this paper.

International Journal of Computer Networks & Communications (IJCNC) Vol.5, No.6, November 2013

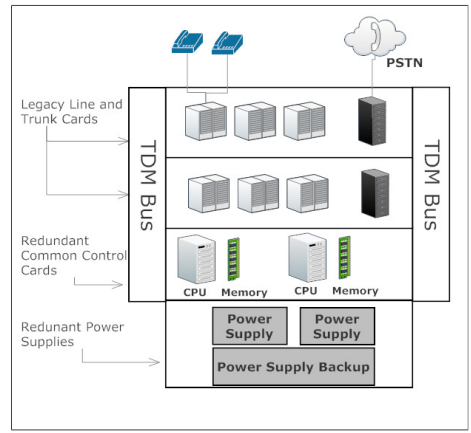

Figure 2: Redundancy in a traditional PBX

The large majority of VoIP developers of in-office telephone solutions have adopted essentially the same legacy centralized philosophy of the traditional PBX and have built what may be termed IP-PBX's in which the RJ-11 from the PBX are replaced by RJ-45 Ethernet wires connected to IP telephones. In this paper we argue that this legacy centralized model grossly limits the exploitation of the full potential of Internet-inspired telecommunications.

In what follows we first describe some approaches that have been adopted for building IP-PBX's and then we describe a distributed IPTS that moves away from the legacy centralized model that allows the exploitation of Internet technologies for converged Voice, Video, Text and Data as well as mobility.

The rest of the paper is organized as follows: Section 2 describes some proprietary and open-source IP-PBX approaches while Section 3 gives an overview of the philosophy of the distributed IPTS motivated in this paper. Section 4 explains Session Initiation Protocol, which is at the core of the message exchange of the IPTS, and Section 5 gives a more detailed description of a prototype of a completely functional and distributed IPTS. In Section 6, the IPTS functionality and preliminary feature set are explained and Section 7 gives concluding comments and direction for future work.

## 2. THE IP-PBX APPROACH

### 2.1. Cisco's Solution

A Cisco IP-PBX system developed on traditional PBX architecture is shown in Figure 3 [5]. Despite claims of high reliability, the drawback is its centralized design, which suffers from all the negative attributes of traditional PBX design and is only a small step in advancement for VoIP. The cost of Cisco's system is high; a BTS Feature License for 6.0.3 Release SIP Update feature costs $4,200,000, not including the primary and backup hardware costs or IP telephone costs. [6].

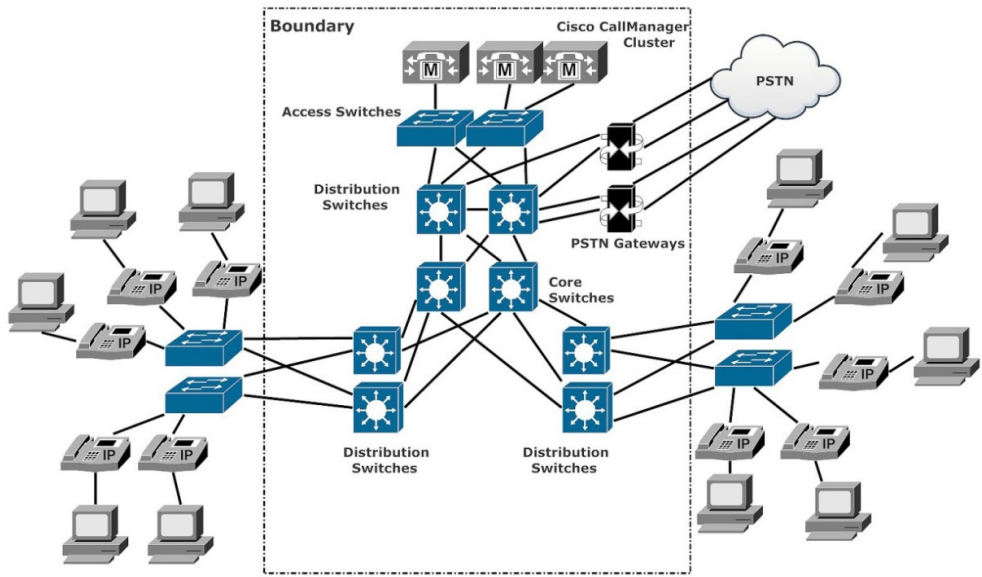

Figure 3: Cisco's Five-Nines System [5]

## 2.2. Shoretel System

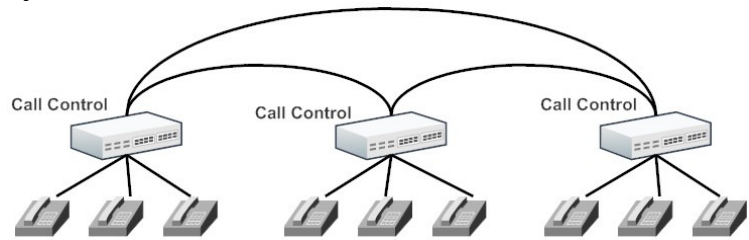

Figure 4: Distributed Call Control [8]

Shoretel's distributed VoIP system (Figure 4), as described in their white paper [8], utilizes a distributed call control component and employs modules with relatively smaller number of components, thus making it less vulnerable to failures. Shoretel claims their modular design makes the components easier and cheaper to repair, but still uses an interconnected set of centralized systems.

Shoretel uses an N+1 redundancy model, which removes the possibility of a single point of failure, for their system. In this method of redundancy, instead of a central component to manage the entire system, the load is distributed amongst multiple units. If the number of units is n, then in case of a fault in any one unit, only 1/*nth* of the users are affected. If another unit is added, then the system becomes n+1 redundant, which becomes part of the same system. In case one of the units fails, the load is distributed amongst the rest. The distributed call control component increases the network availability because all the processing of calls is handled by the nearest call control component. In the case of WAN failure, the individual subcomponents are self-sufficient. In the case of a hardware or network failure, separate failover trunks can do the job of routing the calls. There are also provisions for analog trunks in case the digital trunks fail. Another feature to protect the system from network failures is its ability to use the PSTN even for internal calls. Therefore, even if the IP network fails, the internal calls are routed out to the PSTN, and then in turn to the destination using Direct Inward Dialing (DID).  Software reliability was not available

124

at the time of its publishing; however the average system base cost is about $3000, not including the IP telephones [9].

### 2.3. Open-source Based IP PBX Systems

Several open-source projects have developed technologies that are used to build centralized IP-PBX's - though the same products can be used as building blocks in a non-centralized IP telephony system, as motivated in this paper. For example, Asterisk [10] and FreeSWITCH [11] are commonly used to build IP-PBX systems that are centralized.

## 3. THE IP-BASED TELECOMMUNICATION SYSTEM APPROACH

Instead of developing a central IP-PBX, an IP-based Telecommunication System (IPTS) as described in this paper uses a highly reliable and well designed LAN connection as the backbone (or essentially the backplane – to use PBX/IP-PBX analogies) and a fully distributed and Internet-inspired design to provide in-office IP-based local and long-distance communication. The voice data can be compressed, and redundancy in the voice samples can be removed before they are buffered at the IP gateway for transmission, leading to a very efficient use of bandwidth. The use of IP-based network also makes it extremely convenient to network multiple IPTS's simply and inexpensively to create a distributed regional or global network, without the requirement of special long distance trunks.

Perhaps the biggest advantage of the IPTS design is the flexibility that they provide and this is leveraged to implement a fundamental paradigm shift in design philosophy. Instead of using central, expensive components at the core that drive the entire system, this design uses many inexpensive, replaceable components that can be upgraded, replaced, or removed without disrupting the operation of the live IPTS system. The end devices are intelligent IP phones or softphones that manage their own presence and availability in the network. They also control parameters like call forwarding rules and voicemail time-outs among other settings. Most of the updates and enhancements to the server are software based. The administrator can access the server remotely through the Internet or another IP-based network. Changes to the system, like modification of particular parameters, addition of new subscribers, or changing the privileges of existing subscribers can be made extremely conveniently. Upgrading the server is also usually a software operation and hence proves to be inexpensive. IP-based systems are also highly scalable because of their modular design. Extra modules can be added at reasonable costs. The users also benefit from the portability that the IP-based network provides. By using a softphone, they can access the telephone network from anywhere in the world where an Internet connection is available. All that is required is the IP address of the server and the authentication information of the user.

The goal of this research is to provide a converged distributed system that provides phone calls, text, email, and video routed to a single device while on-the-go, with "follow-me" services, for home and office calls.

The innovative system design is a feature-rich, open source, and fully functional IPTS developed in a manner that is well suited for a distributed design. It is managed by a Session Initiation Protocol (SIP) router, called Kamailio, running on Alpine Linux, with additional features like music on hold, conference call, automated attendant and voicemail. The media functions are provided by FreeSWITCH, which can also act as the Session Border Controller (SBC).

In the next section we give a brief overview of the SIP protocol on which the messaging system in the IPTS is based.

## 4. SESSION INITIATION PROTOCOL

The Session Initiation Protocol (SIP) is one of the most popular protocols used for setting up VoIP calls and is the crux of the IPTS. It is responsible for initializing, modifying and tearing down sessions [12]. The addressing for these sessions are based on Uniform Resource Identifiers (URI) of the involved parties and not the terminals that they are using.

SIP is a text based application-layer protocol, and its syntax is very similar to the Hypertext Transfer Protocol (HTTP). It does not serve as a media gateway and is solely responsible for the session setup/tear-down signaling. SIP does not define the media transfer protocol; it can be used over either TCP or UDP, and by default uses port number 5060. The similarity of SIP to HTTP allows compatibility with web browsers. The SIP message can be of any format, so various kinds of information may be transmitted via SIP. It may contain messages from other protocols such as Real Time Protocol (RTP), Session Description Protocol (SDP), Resource Reservation Protocol (RSVP) and Real Time Streaming Protocol (RTSP).

SIP is responsible for determining the location of the end point to be used based on the URI, with the help of a DNS server and intermediary proxies. Availability of users and their willingness to establish the communication link is negotiated before a call is established and prior to the flow of information/media. Call initiations, transfers, holds, and session termination are all managed by SIP.

A SIP server accepts requests from a User Agent Client (UAC) and sends back responses. The server may act as a proxy server, in which case it can act as a client and forward requests to another server on behalf of a client. The server also functions as a registrar, accepting REGISTER requests, and checking if the UAC is authorized to register with the network. The user can only make a call through a SIP proxy if he/she is registered. The SIP Proxy server forms a triangular topology with the user agent server and client as shown in Figure 5 [12]. The proxy server receives requests from the UAC, and decides where to forward that request. It may either forward it to a User Agent Server (UAS) or to another proxy. The response also follows the same path in reverse. In case the server finds multiple destinations for the requests, it can fork the request and send it to all of them.

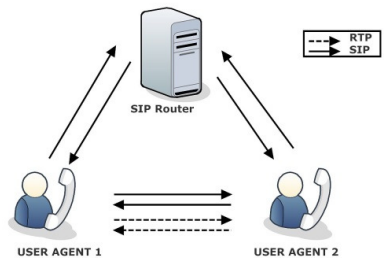

Figure 5: Triangular Topology

The basic SIP protocol [13] [14] uses basic six message types, or methods, which take care of all the functions described in the previous section. These are:

- INVITE: This is a request sent by a UAC to establish a call with another UA. It includes the parameters and description of the media in the SDP format.

126

- ACK: Used to acknowledge the receipt of the response to an INVITE message. It is the last message send during the initial establishment of the call.
- OPTIONS: Used to query about the capability of another user agent.
- BYE: Sent when a user agent wishes to terminate the call.
- CANCEL: Used to cancel a pending request.
- REGISTER: Used to register to a SIP server as explained earlier.

The requests listed above generally result in a response code. The different response codes are shown in Table I [12].

TABLE I. SIP RESPONSE CODES

| Response Code | Name | Meaning |
| --- | --- | --- |
| 1xx | Provisional | Request received, processing |
| 2xx | Success | Action successful (acts as ACK) |
| 3xx | Redirection | Further action required |
| 4xx | Client Error | Current server cannot process |
| 5xx | Server Error | Server failed to process request |
| 6xx | Global Failure | No server can process request |

## 5. COMPONENTS OF THE IP-BASED TELECOMMUNICATION SYSTEM

### 5.1. Overview

A fully functional IPTS consists of several components responsible for various aspects of the communication, Figure 7. A generic overview of each of these components and the role they play in the system is now given [15]. This is followed by a detailed description of the actual components used and roles employed in the system.

1) *SIP Proxy:* This is the central component of the system, which in this set-up is Kamailio. It is responsible for registering users and maintaining a user database. It also takes care of setting up and tearing down of VoIP connections. It does not handle the actual multimedia traffic.
2) *User Administration and Provision Portal:* The administration portal allows users to manage their subscription, and the administrator to regulate the subscribers' privileges.
3) *PSTN Gateway:* The PSTN gateway enables VoIP subscribers to make calls to PSTN subscribers.
4) *Media Server:* A media server is required to deal with functions that require interactive media communication between the IPTS system and the user agents.
5) *Media Proxy:* Directing outgoing calls to the network of the VoIP provider requires the resolving of certain issues. One of the main concerns is NAT traversal. Detail of this functionality is provided in section 8.4.
6) *Monitoring Tools:* These tools are required to debug any problem in the SIP server. These may include a protocol analyzer, and packet sniffing tools like ngrep, Wireshark, tcpdump and ethereal. Kamailio has a module called SIP Trace, which is also useful in detecting errors in the SIP server's operation.

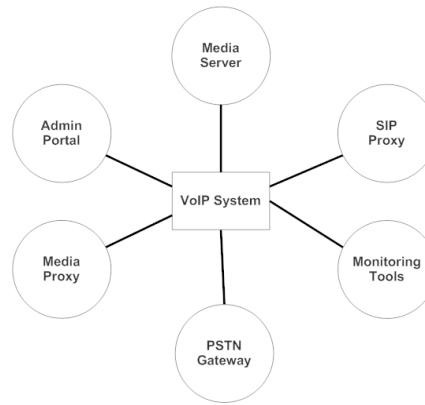

Figure 7. VOIP System Components

The prototype IPTS was built using Kamailio, a Session Initiation Protocol (SIP) router for the setting up and routing of calls and FreeSWITCH, which acts as the media server to provide extra features such as music-on-hold, voicemail, automated attendant, conference calling etc. Calls to the Public Switched Telephone Network (PSTN) are supported with the help of FreeSWITCH, which also functions as the Session Border Controller (SBC) and provides a SIP-based trunk to the service provider. Alpine Linux is the operating system used for implementing the Kamailio and FreeSWITCH since it is highly suitable for running IP-based telecommunication systems, because of its security, speed and run-from-RAM capabilities.

The individual components of the system are described in more detail in the following sections.

### 5.2. Alpine Linux

The operating system (OS) on which the IPTS has been set up is Alpine Linux. It is a Linux operating system specially designed for routers, firewalls, Virtual Private Networks (VPNs) and VOIP [16]. It is very simple to install and can run on a USB stick, which is how we are running it. It is intended for the experienced Linux user and gives the user full control over the system. Alpine Linux uses PaX security, which implements least privilege protection for memory access. This creates a highly secure kernel which is extremely difficult to penetrate, protecting it from bugs [17].

Alpine Linux takes about 4-5 MB of space, excluding the kernel, as compared to gigabytes in the case of most operating systems. It can run from RAM, reducing risk caused by hard disk failure. For RAM based installs, data can be backed up using the Linux Backup Utility (LBU). LBU saves all the configuration information in a single file, which can then be used when the system is restarted to get back the previous saved configuration. The configuration file can also be copied to any compatible Alpine Linux system, resulting in the exact same settings on a separate USB stick. Therefore, the OS is highly portable.

A very useful feature in Alpine Linux is the Alpine Configuration Framework (ACF), a web configuration utility. ACF thus gives the user a GUI for configuring nearly all the settings.

### 5.3. Kamailio

Kamailio is the open source SIP router used in the IPTS. It can handle thousands of calls per second even on low-budget hardware. It also acts as the registrar. IP versions 4 and 6 both are supported by Kamailio, and it has support for communication via TCP, UDP, Transport Layer Security (TLS) and Stream Control Transmission Protocol (SCTP) [18].

Kamailio has been written entirely in the C language. This makes it extremely portable and expandable. New modules for any specific purpose can be easily written using the C language [17]. It is also possible to access scripts written Lua Python, Perl or other supported languages from inside Kamailio, which gives the administrator complete control of the operations of the SIP server.

Kamailio is capable of operating in stateless and stateful modes and provides NAT traversal support for SIP and RTP traffic. It also has features for routing fail-over and replication for high availability. This feature is highly suited to our goal.

The accounting of the calls through Kamailio is event based, and the parameters of the accounting are easily configurable. It also has features for accounting of multi-leg calls. The data of the users is usually stored in a database, for which we have used PostgreSQL as the database management system [17]. All the necessary information about the subscribers is stored in the PostgreSQL database.

The inter-connection of Kamailio to other services like Media Servers and PSTN gateways is fairly simple. Creating a fully functional IPTS with Kamailio as the SIP proxy is thus easily achievable.

1) *Scalability:* When operated on systems with 4GB memory, it can serve about 300,000 online subscribers at one time, and when used as a load balancer in stateless mode, it achieves rates of over 5000 call setups per second [18].
2) *Kamailio Architecture:* Kamailio consists of a core around which all the features described above are built. The core handles the basic functionality as a SIP server and Registrar. The majority of functions are handled by parts of the architecture known as modules. Their functionality is added through new commands and parameters used inside scripts. Each module usually takes care of one feature and is independent of the other modules. At the time of writing, almost 200 modules exist and the ones that are needed by the system are 'loaded' into the configuration file [19]. New modules can be easily added without affecting the core or any other module.
3) *The Configuration File:* The kamailio.cfg configuration file is the place where the entire IPTS system is shaped. A large number of parameters are available through the configuration file. All the routing criteria are described here. Almost every scenario of the SIP protocol can be defined along with the actions needed. The file allows the administrator to access and change parameters specific to a user or a call, which gives him/her total control over the management of the IPTS system. In addition to this, some changes may also be made during run-time.
4) *Role of Kamailio in the IPTS:* Kamailio is the most significant component of the IPTS as it performs the role of the 'core' of the network, much like backbone routers form the 'core' of the Internet. Like backbone routers, the IPTS allows for multiple routers (pathways to UAC's) providing high availability in case of component failures. In many implementations Kamailio also performs the duties of the SIP registrar. In the IPTS a PostgreSQL database of the registered users is maintained, and Kamailio verifies the authenticity of the users based on this database.
Once registered, users can make calls according to the privileges provided per their subscription and agreement with the service provider. Call origination and setup will go to the Kamailio server, which then determines the type of call requested and performs authentication checks. Based on the type of call, the routing decision is then made. Calls may be immediately routed to the destination if it is an internal call, or a DNS query may be performed to figure out the correct destination.

During regular operation, Kamailio may run in stateless mode, which means that once calls are setup and established Kamailio serves as a simple packet forwarder. The performance of the system is improved by reducing the load on the SIP router, enabling it to support more users simultaneously, since Kamailio discards the messages immediately after forwarding.

Additionally, Kamailio can also perform the accounting and billing of the calls once they are concluded. This operation takes place when one of the parties involved in the call hangs up, thereby sending a BYE message to the SIP server. This function may require the SIP router to function in a stateful mode.

## 5.4. FreeSWITCH

FreeSWITCH is an open source IP telephony platform. It is highly scalable and works with a comprehensive range of communication protocols. It can be used as a simple switching engine, a PBX, a media gateway, or a media server to host IVR applications [20]. It makes complicated functions like voicemail and music on hold fairly simple to incorporate into the phone system. In the designed IPTS telephone system, FreeSWITCH functions as a media server and an SBC.

The design of FreeSWITCH, like Kamailio, is based on a stable central core with modules for specific functionality. The core provides interfaces to these modules, allowing the developers to manage the system efficiently, and if needed, add their own modules conveniently. The various modules that operate on the core are completely independent of each other. Each module generates a number of generic events, which enter into the core. The other modules are programmed to listen for these events and are driven by them accordingly.

The most important modules in the FreeSWITCH architecture are the Endpoint modules. These modules provide an interface to one of the supported communication protocols. Such a connection between FreeSWITCH and a protocol such as SIP or H.323 is known as a session. The connection is based on the parameters and settings defined in the XML configuration files.

The basic operation of FreeSWITCH is the following. First, a SIP phone sends call setup message to the SIP module. The call is then passed to the core state machine. This brings FreeSWITCH in the routing state. It now looks for a module called the Dialplan, which is an XML file describing call specific actions for each scenario, known as an extension. The Dialplan now builds a task list for the current call and inserts instructions into the session object. After this, FreeSWITCH goes into the execute state, sequentially implementing the tasks described in the Dialplan. The execution is done by the Application module. The task list is created by the Dialplan in the form of the application name and the arguments it requires. The applications loaded on the core will be listening for these tasks and will execute the ones intended for them.

The configuration can be written in a modular fashion, with different XML files for different purposes. The files can then be included in the central configuration file, where the processing of each call starts. This is particularly useful while creating user accounts. Each user account can be created as a separate file, therefore making it extremely convenient to manage the privileges of the individual users. The users may also be grouped together and have a common account file.

A session border controller (SBC) is a device located at the logical boundary of two networks, which controls the communication between them. It resolves compatibility issues arising as a result of differences in the administration and protocols of the bordering networks, providing interoperability in spite of the mismatch. It can also provide security features, keeping a check on the volume of traffic entering the network, and hiding its topology from surrounding networks. Some of the functions an SBC can perform are listed below [21]:

- Network topology hiding
- Protection from intentional and unintentional flooding of a network
- Protection from unauthorized access to a network
- Resolving Network Address Translation (NAT) issues
- Protocol conversion
- Transcoding of media traffic
- Translation of phone number formats
- Providing additional QoS support

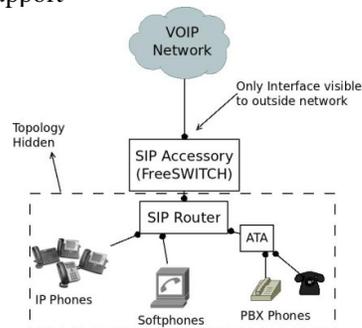

Figure 8. FreeSWITCH as SBC

FreeSWITCH also serves as the SBC for the IPTS phone system, where it is used for network topology hiding and to resolve NAT traversal issues. FreeSWITCH is used to access the external networks, such as the PSTN and the network of the VoIP provider. This can be seen in Figure 8. It thus acts as the single interface visible to these external networks, hiding the SIP proxy and the rest of the topology. This makes communication simpler, and also makes the system less vulnerable to attacks. In order to perform this function, FreeSWITCH acts as a Back-to-back User Agent (B2BUA). This means that it answers requests from the IP phones, and then generates a new SIP call to the destination. The difference between a B2BUA and a SIP proxy is that all the data, including the media streams passes through the B2BUA. Hence it stays in the loop of the communication for the entire duration of the call. This is shown in Figure 9.

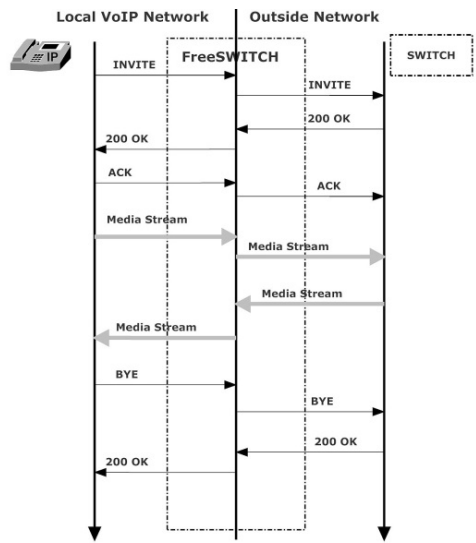

Figure 9. FreeSWITCH as B2BUA

The system created currently resides in a private network, with private IP addresses. This creates a problem, because a NAT device can only convert IP addresses at Layer 3 and not on the Application layer [21]. This culminates in a situation where the call is connected, but audio data is not able to traverse the network. FreeSWITCH resolves this issue by staying in the loop of the communication and changing the private IP address of the SIP devices to its own public IP address before relaying it to the outside world, thus acting once again as a B2BUA.

## 6. OPERATION OF THE COMPLETE SYSTEM

### 6.1. Overview

The IPTS offers a novel approach to IP telephony. It provides all the features present in existing IP-PBXs, but at much lower costs as it is based entirely on open source software and has very modest hardware requirements. It runs on Random Access Memory (RAM), which makes it extremely easy to maintain and highly portable system. The USB disks running the system can be plugged into any computer with the appropriate amount of RAM and an Internet connection, resulting in the exact same functionality. The USB is made un-writable during normal operation, which leads to a highly protected tamper-proof system. Duplicating the USB disks is fairly straightforward, a feature that is beneficial for backing up the system as well as expanding the system.

The prototype system is Internet-inspired, and the intelligence of the network lies at the terminal equipment, as mentioned in Section 5. The server does not impose any form of features on the phones. This allows the user flexibility to tweak system parameters by means of their SIP phones, irrespective of the type of end system.

It is a distributed type of system, in the sense, that the SIP server is not in the communication loop at all times. It is only involved in the setting up of the call. The media does not traverse through the SIP server; it goes directly from one end system to the other. This reduces the traffic flow through the server making it much more efficient. As a result, the number of dropped calls is reduced. It also allows the server to support more calls simultaneously.

International Journal of Computer Networks & Communications (IJCNC) Vol.5, No.6, November 2013

The distributed and end system oriented design of the IPTS phone system also gives it structure, which centralized servers lack. Different kinds of functions are divided in an organized manner across the system, instead of one server doing it all. For instance, Kamailio is responsible for the call set up and FreeSWITCH handles the media functions and also acts as the Back to Back User Agent (B2BUA). IP phones are able to handle call transfers and call forwarding. They can also control parameters of other functions such as the ringer timeout. The structure of the system is depicted in Figure 10 in contrast to traditional phone systems. As we can see, the system will connect to the PSTN network as well and can support Direct Inward Dialing (DID).

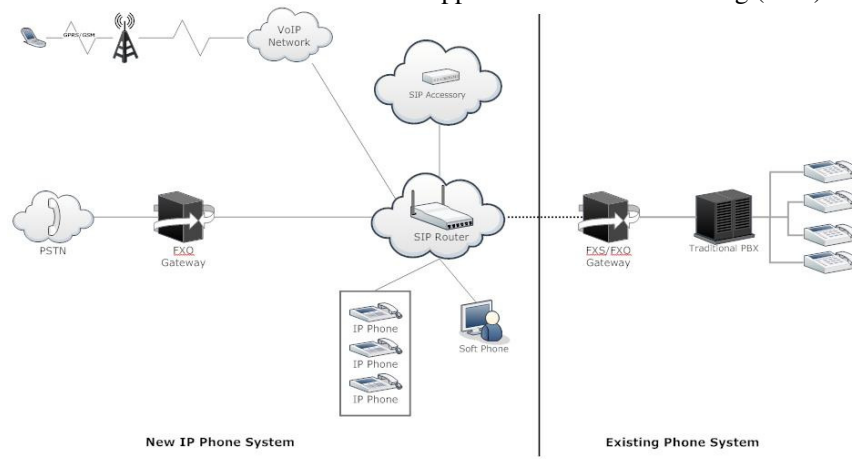

Figure 10. Complete IPTS

The system provides several extra functions apart from basic voice calls. The functions that we are concentrating on include:

- Internal Calls
- External Calls
- Music on Hold
- Voicemail
- Conference Calls
- Automated Attendant

## 6.2. Call Processing

As described earlier, the information and credentials of the registered users are stored in a PostgreSQL database. Users may be added manually by extension number or provisioning. In the case of provisioning, once a registered device comes online, it gets all the configuration information by sending a DHCPREQUEST to the DHCP server with option 66, followed by a request to the provisioning server.

Every call, irrespective of its destination goes through some initial steps before it can finally be routed according to the destination URI and other factors, such as the availability of the called party. When a user dials an extension or external number, the SIP enabled phone generates an INVITE message with the appropriate destination URI. This message is sent to the SIP router, Kamailio. The message is first checked for corrupt or illegal fields, such as the length of the message being too large, or the number of hops allowed being too many. Now, based on the destination URI, a number of actions may be taken by Kamailio. These are now described:

1) *Internal Calls:* An internal call is defined as a call whose source and destination are both registered to the same SIP server. These calls are handled only by Kamailio. Once the initial processing of the INVITE message is completed; if the destination is found in the table, Kamailio concludes that this is an internal call and accordingly sets the parameters of the call. The INVITE message is then relayed to the destination extension. The destination SIP-enabled telephone receives this INVITE message, and if not busy, replies with a 200 OK message. The 200 OK message reaches the Kamailio server and is relayed to the caller. Once the caller receives 200 OK, it replies with an ACK, and the call is established. The transfer of voice data is not handled by SIP. When one of the participants of the call hangs up, a BYE message is generated and relayed to the other participant through the Kamailio server. The call ends when the message is acknowledged. After the call ends any required post processing like billing may be performed.
2) *External Calls*: When a registered user on the system dials a number outside the network, it is known as an external call. To be able to make external calls we need to have a VoIP account with a service provider. These calls are routed through FreeSWITCH. The calls are identified by first dialing a "9" before the telephone number. The call initially goes through the same steps as the internal calls. However, in the step where Kamailio looks for the destination in the registered users table, it fails. Kamailio then identifies with the help of the leading 9 that it is an external call. At this point, it redirects the call to FreeSWITCH for further routing. Within the FreeSWITCH configuration files, an external SIP profile is defined, which has the credentials of the VoIP account which will allows us to make the external call. In the FreeSWITCH dialplan, a condition is defined for destination phone numbers starting with 9. The call is bridged to the VoIP service provider's server using the mod_sofia module. The authentication is performed using the information provided in the external SIP profile. Hence, the call is established. FreeSWITCH now stays in the call for the entire duration, acting as a B2BUA.
3) *Music on Hold:* The phone system provides music on hold (MOH). This means that whenever a user registered on the server puts a call on hold the other participant of the call gets music streamed to his/her phone. This requires a media proxy, which is FreeSWITCH in this system. A Music on Hold extension is defined in the FreeSWITCH dialplan. When a call is placed on hold, a new branch is formed by Kamailio, and the call is redirected to the MOH extension. The pressing of the hold button culminates in a new INVITE being generated by the user agent with a blank payload.
4) *Voicemail:* If a user does not answer a call, it fails over to voicemail, where a message can be recorded. Every registered user requires a FreeSWITCH voicemail account to be able to receive messages. This can be done using Alpine Linux's web configuration utility or by creating individual configuration files for each of the users in the FreeSWITCH directory. There is an entry for the voicemail extension in the FreeSWITCH dialplan that uses the voicemail application. The messages can also be forwarded to an e-mail address if needed.
5) *Conference Calls:* The conference call feature allows more than two users to talk to each other simultaneously. A series of extension numbers are reserved for conference calls. For example, in this system, the extension numbers of the form 30XX are used for conference calls. This means that a conference call can be initiated by a user simply by dialing any extension of this form. Other users who wish to join the conference call can do so by dialing the same extension. The behavior of FreeSWITCH when a conference call is initiated is similar to when Music on Hold has to be played. The conference call is handled by the application called conference. If this application is initialized correctly with the correct domain address, it takes care of all the bridging expected.
6) *Automated Attendant:* This feature provides the caller with an Interactive Voice Response (IVR) system. This can be used to redirect incoming calls without an operator. The call is redirected through FreeSWITCH in a similar manner as the conference calls.

## 7. CONCLUSION

A fully functional distributed IP-based Telecommunications System was designed and implemented. It provides the common features of a traditional office PBX, namely music-on-hold, voicemail, automated attendant and conference calls. External calls to any PSTN extension can be made by redirecting the call to FreeSWITCH and using it as a B2BUA. Intelligent SIP phones used in the system can directly take the configuration settings from the provisioning server, and get automatically registered. The phones can handle call transfers on their own. The user can also modify any of the settings as per his/her requirement. The phone will then inform the provisioning server of these changes. The designed system is thus highly flexible.

The system has been created in a manner that highly suits the design of a distributed IP telephony system. As a result, it is the first step in achieving a single device for all communication needs. Future work on the system will be on improving reliability by adding redundant components and also automatic provisioning, with an examination of the relationship of the approach articulated in this paper with emerging the IP Multimedia Subsystem principles [22].

[1] Vonage Home Page. [Online]. Available: http://www.vonage.com/
[2] Magic Jack. [Online]. Available: http://www.magicjack.com/plus-v05/
[3] AT&T International Rates [Online]. Available: http://www.att.com/media/att/2013/support/pdf/uverse_international_rates.pdf
[4] "Telephone Communications: Bypass of the Local Telephone Companies", RCED-86-66, Aug 18, 1986 [Online]. Available: http://www.gao.gov/products/RCED-86-66
[5] A. Leon-Garcia, "Communication Networks", 2nd Ed. New York, NY, USA: McGraw-Hill, 2004.
[6] "IP telephony: The five nines story," White Paper, Cisco Systems, 2002. [Online]. Available: http://www.cisco.com/web/IT/solutions/pdf/ipcom/5nine_story_wp.pdf
[7] "Cisco FL VOIP Price List - Industry Solutions - Cisco Systems" [Online]. Available: http://www.cisco.com/web/strategy/government/fl_voip/price_list_archive.html
[8] "IP telephony: Reliability you can count on," White Paper, Shorten, 2009. " [Online]. Available: http://www.shoretel.com/resource_center/results?#/?white_paper=&content=92715959
[9] "ShoreTel Pricelist – Peppm" " [Online]. Available: http://www.peppm.org/Products/ShoreTel/price.pdf
[10] Asterisk Home Page " [Online]. Available: http://www.asterisk.org/
[11] FreeSWITCH Home Page " [Online]. Available: http://www.freeswitch.org/
[12] J. Rosenberg, H. Schulzrinne, G. Camarillo, A. Johnston, J. Peterson, R. Sparks, M. Handley, and E. Schooler, "SIP: Session initiation protocol," IETF RFC 3261, Jun. 2002. [Online]. Available: http://datatracker.ietf.org/doc/rfc3261/
[13] J. Rosenberg, H. Schulzrinne, "Reliability of Provisional Responses in the Session Initiation Protocol (SIP)," IETF RFC 3262, Jun. 2002. [Online]. Available: http://www.ietf.org/rfc/rfc3262.txt
[14] C. Holmberg, E. Burger, H. Kaplan, "Session Initiation Protocol (SIP) INFO Method and Package Framework," IETF RFC 6086, Jun. 2011. [Online]. Available: http://tools.ietf.org/html/rfc6086
[15] F. E. Goncalvez" Building Telephony Systems with OpenSER: A step-by-step guide to building a high performance Telephony System". Birmingham, UK: Packt Pub., 2008.
[16] About alpine Linux. [Online]. Available: http://alpinelinux.org/about
[17] H. A. Latchman, N. Angelacos, and N. Copa, "Enterprise VOIP solutions with alpine Linux," in Slashroots 2011 Developer's Conference.
[18] Kamailio official online documentation. [Online]. Available: http://www.kamailio.org/docs/
[19] "Kamailio Modules - v4.0.x (stable)". [Online]. Available: http://kamailio.org/docs/modules/stable/
[20] A. Minessale and M. S. Collins, "FreeSWITCH 1.0.6." Birmingham, UK: Packt Publishing, 2010.
[21] P. Park, "Voice over IP Security". Indianapolis, IN, USA: Cisco Press, 2009.

**Authors**

**Carlton Thompson** Completed BS & MS degree at the University of Florida in the Department of Electrical Engineering. Previous research was done using Electrical Impedance Tomography (EIT) for the development of a system for capturing and reconstructing impedance images of the head, and extracting information about bleeding-related impedance changes from background noise. Results gathered from two and three-dimensional finite element models of the head will be used to test reconstruction methods on a skull model. Current work is on development of a completely distributed IP-based Telecommunications System (IPTS). 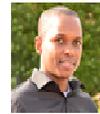

**Dr. Haniph A. Latchman** is a Rhodes Scholar and received his Ph.D. from Oxford University in 1986 and his Bachelor of Science degree (First Class Honors) from the University of The West Indies-Trinidad and Tobago, in 1981. Dr. Latchman joined the University of Florida in 1986 and is presently Professor of Electrical and Computer Engineering. Dr. Latchman has received numerous teaching and research awards, including the 1998 University of Florida Teacher of the Year and the 2000 IEEE Undergraduate Teaching Award. Dr. Latchman is the author of more than 180 Journal and refereed conference papers and 4 books. He has directed 24 Ph.D. Dissertations and 35 MS Theses. He has also served as Associate editor and guest editor for several leading Journals in his research areas of Robust Control Systems and Communication Networks. For further information about Dr. Latchman's research and teaching program, visit http://www.list.ufl.edu. 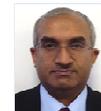

**Nathan Angelacos** has worked with various networking technologies for 25 years. Most projects involved the impossible - like building a wireless mesh network blocks away from NYC's Office of Emergency Management transmitters. He has done large-scale SIP / VoIP deployments in a variety of topologies, including VPN and DMVPN networks. He is currently the senior systems engineer for a SIP wholesale provider. 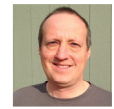

**Bharath Kumar Pareek** received B.E. in Electronics and Communications from Visvesvaraya Technological University, India in 2011 and received his M.S. in Electrical and Computer Engineering with Computer Engineering as his specialization from University of Florida in 2013. He is currently a Research Assistant at Laboratory for Information Systems and Telecommunications, Department of ECE, University of Florida. His research interests include Voice over Internet Protocol, Machine Learning, Cloud Computing and Computer Networks. 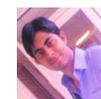